\documentclass[final,5p,times,twocolumn,nofootinbib,nopreprintline]{elsarticle}
\usepackage{amsmath,amssymb,amsfonts}
\usepackage{graphicx}
\usepackage{hyperref}
\usepackage{slashed}
\usepackage{booktabs,tabulary,longtable, multirow}
\usepackage{mciteplus}
\mciteErrorOnUnknownfalse
\usepackage{color}
\usepackage{soul}

\usepackage{bibentry}
\newcommand{\ignore}[1]{}
\newcommand{\nobibentry}[1]{{\let\nocite\ignore\bibentry{#1}}}

\makeatletter
\def\bibinfo@X@title#1,{\ignorespaces}
\makeatother

\def\chizero{\ensuremath{\chi_{c0}(2P)} }
\def\chitwo{\ensuremath{\chi_{c2}(2P)} }
\def\ggdd{\ensuremath{\gamma\gamma\to D\bar{D}} }
\def\jp{\ensuremath{J/\psi}}

\begin{document}

\title{Dispersive analysis of the $\gamma\gamma \to D \bar{D}$ data and the confirmation of the $D \bar{D}$ bound state}
	
\author{Oleksandra Deineka}
\author{Igor Danilkin}
\author{Marc Vanderhaeghen}
\address{Institut f\"ur Kernphysik \& PRISMA$^+$  Cluster of Excellence, Johannes Gutenberg Universit\"at,  D-55099 Mainz, Germany}
	
\date{\today}

\begin{abstract}
In this paper, we present a data-driven analysis of the $\gamma\gamma\to D^+D^-$ and $\gamma\gamma\to D^0\bar{D}^0$ reactions from threshold up to 4.0 GeV in the $D\bar{D}$ invariant mass. For the $S$-wave contribution, we adopt a partial-wave dispersive representation, which is solved using the $N/D$ ansatz. The left-hand cuts are accounted for using the model-independent conformal expansion. The $D$-wave $\chi_{c2}(3930)$ state is described as a Breit-Wigner resonance. The resulting fits are consistent with the data on the invariant mass distribution of the $e^+e^- \to J/\psi D\bar{D}$ process. Performing an analytic continuation to the complex $s$-plane, we find no evidence of a pole corresponding to the broad resonance $X(3860)$ reported by the Belle Collaboration. Instead, we find a clear bound state below the $D\bar{D}$ threshold at $\sqrt{s_B} = 3695(4)$ MeV, confirming the previous phenomenological and lattice predictions. 

\end{abstract}
	
\maketitle

\section{Introduction}\label{sec:intro}
The growing interest in the charmonium mass region in recent years is nourished with new experimental discoveries. Ever since the Belle Collaboration discovered the $X(3872)$ \cite{Belle:2003nnu} extremely close to the $D^0\bar{D}^{*0}$ threshold, a plethora of new states has been observed. Nevertheless, only a few of them are unambiguously identified.  For comprehensive reviews, we refer to \cite{ Chen:2016qju,*Esposito:2016noz,*Olsen:2017bmm,*Guo:2017jvc,*Karliner:2017qhf,*Brambilla:2019esw}. One of the controversial examples is the identification of $\chi_{c0}(2P)$ state. The first attempts of its assignment date back to 2010, when in \cite{Liu:2009fe} it was proposed to identify $\chi_{c0}(2P)$ with the narrow resonance $X(3915)$ seen by the Belle \cite{Belle:2004lle, *Belle:2009and} and BaBar \cite{BaBar:2007vxr, *BaBar:2010wfc} Collaborations. This assignment was later supported by the spin-parity analysis by the BaBar Collaboration \cite{BaBar:2012nxg}. However, as it was pointed out in several other works \cite{Brambilla:2010cs,*Eichten:2005ga,*Guo:2010ak}, $X(3915)$ is a problematic candidate for $\chizero$ due to its narrowness, dominant decay channels (which contradict the expectations for $\chizero$), and the small mass splitting with the well-established  $\chi_{c2}(3930)$.
			
The alternative candidate for $\chizero$ state may have been already observed in the $\ggdd$ process by both Belle \cite{Belle:2005rte} and BaBar \cite{BaBar:2010jfn} Collaborations somewhere in an energy range from the $D\bar{D}$ threshold up to the $\chitwo$ position. In \cite{Guo:2012tv}, $\ggdd$ data were reanalyzed using two Breit-Wigner functions under the assumption, that the invariant mass distribution is dominated by the resonance structures. In other words, it was assumed that the broad bump located around $\sim 3800$ MeV, which was considered to be a background in experimental analyses, may hide the broad resonance. By fixing the mass and the width of $\chitwo$ to its experimental values the fit to data predicted the existence of $\chizero$ with $M_{\chizero} = 3837.6 \pm 11.5$ MeV  and $\Gamma_{\chizero} = 221 \pm 19$ MeV. The later result was reinforced by the Belle Collaboration \cite{Chilikin:2017evr}, which in the analysis of $e^+e^- \to J/\psi D\bar{D}$ data found the new charmonium-like state $X(3860)$, that decays mainly to $D\bar{D}$ channel. With the mass  $3862^{+26 +40}_{-32-13}$ MeV, the width $201^{+154 +88}_{-67 -82}$MeV and $J^{PC} = 0^{++}$ this state is currently included in the PDG (2021)  \cite{ParticleDataGroup:2020ssz} as $\chizero$.
			 
However, it is still an open question of what has been seen in $\ggdd$ and $e^+e^- \to J/\psi D\bar{D}$ processes. First, the statistics of the Belle data \cite{Chilikin:2017evr}  for the $e^+e^- \to J/\psi D\bar{D}$ process is rather low close to the threshold. Second, the proper resonance analysis should account for the S-matrix constraints, unlike simple Breit-Wigner parametrizations. In \cite{Wang:2019evy} a unitary approach based on the Bethe-Salpeter equation was used to describe the Belle data. No peak structure that justifies the claim for the $X(3860)$ state was found. The same observation was made in  \cite{Wang:2020elp} regarding the $\ggdd$ data from the Belle \cite{Uehara:2005qd} and BaBar \cite{Aubert:2010ab} Collaborations. Instead, this analysis suggested that the behaviour around the threshold is consistent with the $D\bar{D}$ dynamics that encodes a bound state, previously predicted in \cite{Gamermann:2006nm}. On another side, the recent coupled-channel $\{D\bar{D},D_s\bar{D}_s\}$ analysis performed on the lattice with $m_\pi=280(3)$ MeV \cite{Prelovsek:2020eiw} suggests the existence of both: a shallow bound state slightly below $D\bar{D}$ threshold and the broad resonance, comparable to $X(3860)$. Moreover, the situation gets more puzzling by the recent LHCb observation \cite{LHCb:2020bls,*LHCb:2020pxc} of two resonances, sitting at the same mass, the $\chi_{c0}(3930)$ and the $\chi_{c2}(3930)$, with widths around 17 MeV and 34 MeV, respectively, and no evidence of the broad $X(3860)$ state.

The present ambiguity regarding the existing data and the character of the structures present in $\gamma\gamma\to D\bar{D}$ cross sections and $e^+e^- \to J/\psi D\bar{D}$ calls for a theoretical approach, which rigorously implements both the unitarity and analyticity constraints and does not make any assumption about underlying $D\bar{D}$ dynamics. Previously, the once-subtracted partial wave dispersion relation was successfully used for the analysis of the $\pi \pi$ and $\pi K$ scattering in \cite{Danilkin:2020pak}. Within this framework, it is straightforward to perform the analytical continuation of the scattering amplitudes to the unphysical regions and identify the positions of the poles and bound states. Therefore, an application of this technique to the $D\bar{D}$ system can shed more light on the nature of the near-threshold enhancements seen in the experiment.
	
This paper is organized as follows. In Sec.~\ref{subsec:nd}, we describe the partial wave dispersive formalism which we adopt for the S-wave in the $D\bar{D}$ system. In Sec.~\ref{subsec:dwave} we present the details of the tensor $\chi_{c2}(3930)$ resonance. We show our numerical results in Sec.~\ref{sec:results}, which included the analysis of the $\ggdd$ data in Sec.~\ref{subsec:ggdd} and a post-diction to the $e^+e^-\to J/\psi D\bar{D}$ process in Sec. \ref{subsec:jpsi}. A summary and outlook are given in Sec.~\ref{sec:conc}.

\section{Formalism}\label{sec:form}
\subsection{$S$-wave amplitudes}\label{subsec:nd}
	
We consider a $2\to 2$ process described by the partial wave (p.w.) amplitudes $t^{(J)}_{I, ab}$, where $ab$ are the coupled-channel indices with $a$ and $b$ standing for the initial and final state, respectively. In this subsection, we focus only on the $S$-wave, with isospin $I = 0$, and therefore will suppress the labels $I, J$. The unitarity condition can be written in the matrix form as
\begin{align}\label{Eq:Unitarity}
    \text{Disc}\,t_{ab}(s)&\equiv\frac{1}{2i}\left(t_{ab}(s+i \epsilon)-t_{ab}(s-i \epsilon)\right)\nonumber\\
    &=\sum_{c} t_{ac}(s)\,\rho_{c}(s)\,t^*_{cb}(s)\,,
\end{align}
where the sum goes over all intermediate states. The phase space factor $\rho_{c}(s)$ in Eq.~(\ref{Eq:Unitarity}) is given by 
\begin{align}\label{Eq:rho}
    \rho_{c}(s)&=\frac{1}{8\pi}\frac{p_{c}(s)}{\sqrt{s}}\,\theta(s-s_{th})\,,
\end{align}
with $p_{c}(s)$ and $s_{th}$ being the center-of-mass three momenta and threshold of the corresponding two-meson system. The unitarity condition guarantees that the partial-wave amplitudes at infinity approach at most constants. In accordance with that, and based on the maximal analyticity assumption \cite{Mandelstam:1958xc,*Mandelstam:1959bc}, we write once-subtracted dispersive representation
\begin{align}\label{DR_1}
    t_{ab}(s)&=t_{ab}(0)+\frac{s}{\pi} \int_{-\infty}^{s_L}\frac{d s'}{s'}\frac{\text{Disc } t_{ab}(s')}{s'-s} + \frac{s}{\pi} \int_{s_{th}}^{\infty}\frac{d s'}{s'}\frac{\text{Disc } t_{ab}(s')}{s'-s}\nonumber\\
    &\equiv U_{ab}(s) + \frac{s}{\pi}\sum_c \int_{s_{th}}^{\infty}\frac{d s'}{s'}\frac{t_{ac}(s')\,\rho_{c}(s')\,t^*_{cb}(s')}{s'-s}\,,
\end{align}
where $s_\text{th}$ is the lowest threshold and $s_L$ is the position of the closest left-hand cut singularity. Our particular choice of the subtraction point at $s=0$ will be discussed later. In the second line of Eq.~(\ref{DR_1}), we combined the subtraction constant with the left-hand cut contributions into the function $U_{ab}(s)$. The solution to (\ref{DR_1}) can be obtained numerically using the $N/D$ ansatz \cite{Chew:1960iv}
\begin{equation}\label{N/D}
		t_{ab}(s)=\sum_c D^{-1}_{ac}(s)\,N_{cb}(s)\,,
\end{equation}
where the contributions of left- and right-hand cuts are separated into $N(s)$ and $D(s)$ functions, respectively. As a consequence of this ansatz, one needs to solve a system of linear integral equations \cite{Luming:1964,* Johnson:1979jy}
\begin{align}\label{N/D_equations}
	N_{ab}(s)&=U_{ab}(s)+ \\
	&\frac{s}{\pi} \sum_{c} \int_{s_{th}}^{\infty}\frac{d s'}{s'}\frac{N_{ac}(s')\,\rho_{c}(s')\,(U_{cb}(s')-U_{cb}(s))}{s'-s}\,, \nonumber\\
	D_{ab}(s)&=\delta_{ab}- \frac{s}{\pi} \int_{s_{th}}^{\infty}\frac{d s'}{s'}\frac{N_{ab}(s')\,\rho_{b}(s')}{s'-s}\,,
\end{align}
where the input of $U_{ab}(s)$ is required for $s>s_{th}$ only. Note also, that we assume that there are no Castillejo-Dalitz-Dyson (CDD) poles \cite{Castillejo:1955ed}.

We aim to extract the $S$-wave photon fusion amplitude $\gamma\gamma \to D\bar{D}$, which is the off-diagonal term of the coupled channel $\{1=\gamma\gamma, 2=D\bar{D}\}$ system. Note that for the $S$-wave, there is only one $\gamma(\lambda_1)\gamma(\lambda_2) \to D\bar{D}$ helicity amplitude with helicities $\lambda_1=\lambda_2=+1$. By neglecting $\gamma\gamma$ intermediate states in the unitary relation $\rho_1 = 0$, and putting $U_{11} = 0$ (which is proportional to $e^4$ and hence suppressed), the coupled-channel $N/D$ solution reduced down to the separate set of integral equations for the hadronic part
\begin{align}\label{N/D_22}
	t_{22}(s)&=N_{22}(s)/D_{22}(s)\,,\\
	N_{22}(s)&=U_{22}(s)+\frac{s}{\pi} \int_{4m_D^2}^{\infty}\frac{d s'}{s'}\frac{N_{22}(s')\,\rho_{1}(s')\,(U_{22}(s')-U_{22}(s))}{s'-s}\,,\nonumber\\
	D_{22}(s)&=1- \frac{s}{\pi} \int_{4m_D^2}^{\infty}\frac{d s'}{s'}\frac{N_{22}(s')\,\rho_{2}(s')}{s'-s}\,,\nonumber
\end{align}
and for the $\gamma\gamma \to D\bar{D}$ part
\begin{align}\label{AMPD}
	t_{12}(s)&=U_{12}(s)+D_{22}^{-1}(s)\left(-\frac{s}{\pi}\int_{4m_D^2}^{\infty}\frac{ds'}{s'}\frac{\text{Disc}(D_{22}(s'))U_{12}(s')}{s'-s}\right)\,.
\end{align}
The latter requires as input the hadronic $D_{22}$ function given in Eq.~(\ref{N/D_22}) as well as the $\gamma\gamma \to D\bar{D}$ left-hand cuts, $U_{12}$. 

For the case when there is no bound state in the system, Eq.~(\ref{AMPD}) can be obtained from writing the once-subtracted dispersion relation for the quantity $\Omega_{22}^{-1} \left(t_{12}-U_{12}\right)$ \cite{Garcia-Martin:2010kyn,*Dai:2014zta,*Dai:2014lza}, where $\Omega_{22}=D_{22}^{-1}$ is the Omn\`es function \cite{Omnes:1958hv,*Muskhelishvili-book}. However, it is important to emphasize that Eqs.~(\ref{N/D_22}) and (\ref{AMPD}) are universal also for the case when there is a bound state in the system. It is straightforward to show that adding a bound state into $U_{ab}(s)$,
\begin{align}\label{UBS}
	\tilde{U}_{ab}(s)= U_{ab}(s)+\frac{s}{s_B}\frac{g_{ab}^2}{s_B-s}
\end{align}
does not change Eqs.~(\ref{N/D_22}) and (\ref{AMPD}) provided that the binding energy $s=s_B$ is determined by 
\begin{align}\label{Eq:D=0}
    D_{22}(s_B)=1- \frac{s_B}{\pi} \int_{4m_D^2}^{\infty}\frac{d s'}{s'}\frac{N_{22}(s')\,\rho_{2}(s')}{s'-s_B}=0\,.
\end{align}
For the case when there is a bound state in the system, Eq.~(\ref{AMPD}) is equivalent to the once-subtracted dispersion relation for the quantity $\Omega_{22}^{-1}\left(t_{12}-\tilde{U}_{12}\right)$, where the Omn\`es  function is now related to the D-function as $\Omega_{22}=\left(\frac{s_B}{s_B-s}\right)\,D_{22}^{-1}$.

To evaluate the dispersion relations in Eqs.~(\ref{N/D_22}) and (\ref{AMPD}), we need to specify the left-hand cuts. For the photon-fusion process $\gamma\gamma \to D\bar{D}$ the left-hand cuts can be well approximated by the exactly calculable Born contribution,
\begin{align}
    U_{12}(s) &= -\frac{2\sqrt{2}\,e^2 m_D^2}{s\,\beta(s)}\log{\frac{1+\beta(s)}{1-\beta(s)}}\,,
    \\
    \beta(s) &\equiv \frac{2\,p(s)}{\sqrt{s}} =
    \sqrt{1-\frac{4\,m_D^2}{s}}\,.\nonumber
\end{align}
Heavier left-hand cuts exchanges start farther away from the physical region and typically suppressed for the S-wave contribution \cite{Danilkin:2018qfn}. Note, that the choice of the subtraction point in Eq.~(\ref{DR_1}) and consequently in Eq.~(\ref{AMPD}) is motivated by the soft-photon theorem \cite{Low:1958sn}, which states that the Born term subtracted photon fusion amplitude must vanish at $s=0$. As for the $D\bar{D}\to D\bar{D}$ left-hand cuts, little is known about them, except their analytic structure in the complex plane. Since we need the input for $U_{22}(s)$ only in the physical region, one can approximate $U_{22}(s)$ by means of a model independent conformal expansion \cite{Gasparyan:2010xz,*Danilkin:2010xd,*Gasparyan:2011yw,*Gasparyan:2012km}
\begin{align}\label{ConfExpansion}
	U_{22}(s)&= \sum_{n=0}^\infty C_{n}\,\xi^n(s)\,,
\end{align}	
where the conformal mapping variable $\xi(s)$
\begin{align}\label{xi}
	\xi(s)&=\frac{\sqrt{s-s_L}-\sqrt{s_E-s_L}}{\sqrt{s-s_L}+\sqrt{s_E-s_L}}\,,
\end{align}
maps the left-hand cut plane $-\infty<s<s_L$ onto the unit circle \cite{Frazer:1961zz}. The position of the closest left-hand cut branching point $s_L=4(m_D^2-m_\pi^2)$ is determined by the $t-$ and $u-$ channel exchange of two pions. The expansion point $s_E$ (at which $\xi(s_E)=0$) is chosen in the middle of the region where we expect the S-wave contribution to dominate 
\begin{equation}\label{Eq:s_E}
	\sqrt{s_E}=\frac{1}{2}\,\left(\sqrt{s_\text{th}}+\sqrt{s_\text{max}}\right)\,,
\end{equation}
with $\sqrt{s_\text{max}} = 3.86$ GeV. We note that, given the form of $\xi(s)$ in Eq.~(\ref{xi}), the series (\ref{ConfExpansion}) truncated at any finite order is bounded asymptotically. This is consistent with the assigned asymptotic behavior of $U_{22}(s)$ in the once-subtracted dispersion relation (\ref{DR_1}). In the next section, we will determine the unknown $C_n$ in Eq.~(\ref{ConfExpansion}) directly from the data.

Hereafter, to distinguish the amplitudes involving photons from the pure hadronic amplitude, for the $\gamma\gamma \to D\bar{D}$ p.w. amplitudes we introduce the notation $h^{(J)}_{I, \lambda_1\lambda_2}(s)$, where $\lambda_{1,2}=\pm1$ are photon helicities, so that \begin{equation}
    h_{0,++}^{(0)}(s)\equiv t_{12}(s)\,.
\end{equation}
While it is natural to associate any resonant structure with the dynamics in the $I=0$ channel, the $I=1$ amplitude does not have known direct channel resonances and we approximate it by the Born amplitude 
\begin{equation}
	h^{(0)}_{1,++}(s) = -\frac{2\,\sqrt{2}\,e^2 m_D^2}{s\,\beta(s)}\log{\frac{1+\beta(s)}{1-\beta(s)}}\,.
\end{equation}
We note, however, that taking into account the $I=1$ contribution is absolutely necessary to obtain nonequal cross sections for the $\gamma\gamma \to D^+D^-$ and $\gamma\gamma \to D^0\bar{D}^0$ channels.

\subsection{D-wave amplitudes}\label{subsec:dwave}
For the D-wave in the $\gamma\gamma \to D\bar{D}$ process we take into account only the contribution from the isoscalar $\chi_{c2}(3930)$ resonance, which is a radially excited $P$-wave charmonium state. We approximate it by a simple Breit–Wigner form, similar to how it was done for 
$f_2(1270)$ in the $\gamma\gamma\to\pi\pi$ process in \cite{Drechsel:1999rf, *Hoferichter:2011wk} and for $a_2(1320)$ in the $\gamma\gamma\to\pi^0\eta$ process in \cite{Danilkin:2017lyn, *Deineka:2018nuh}. Is it based on the effective Lagrangians of the following form
\begin{align}\label{Eq:Lagrangians}
    {\cal L}_{R\gamma\gamma}=e^2\,g_{R\gamma\gamma}\,\Phi_{\mu\nu}\,F^{\mu\lambda}\,F^{~~\nu}_{\lambda}\,,\nonumber\\
    {\cal L}_{RD\bar{D}}=g_{RD\bar{D}}\,\Phi^{\mu\nu}\,\partial_\mu D\,\partial_\nu D\,,
\end{align}
where $F^{\mu\nu}$ is an electromagnetic tensor and $\Phi^{\mu\nu}$ is a massive spin-2 field. In the first line of Eq.(\ref{Eq:Lagrangians}) it is assumed that the $\chi_{c2}(3930)$ resonance is predominantly produced in a state with helicity-2. The D-wave amplitude is then given by
\begin{equation}\label{pw_D}
	h^{(2)}_{0,+-}(s) = -  \frac{e^2\,g_{R\gamma\gamma}\,g_{R D \bar{D}}}{10\sqrt{6}}\frac{s^2\,\beta^2(s)}{s-M_R^2 + i\,M_R\,\Gamma_R(s)}\, ,
\end{equation}
where $g_{R\gamma\gamma}, g_{R D \bar{D}}$  denote $\chi_{c2}(3930)$ couplings to $\gamma\gamma$ and $D\bar{D}$ channels, respectively.  The $s$-dependent decay width of the resonance we parametrise as \cite{Belle:2009xpa}
\begin{equation}\label{Gamma_D}
	\Gamma(s) = \Gamma_{R} \left(\frac{p(s)}{p(M_R^2)}\right)^5\,,
\end{equation}
with $\Gamma_{R}$ being the width of the resonance at rest. Note, that for simplicity we have not included Blatt-Weisskopf factors in Eqs.~(\ref{pw_D}) and (\ref{Gamma_D}), which only slightly change the cross section in the considered region but introduce additional dependence on the unknown interaction radius, which cannot be fixed given the quality of the present data. While for the mass and the width of $\chi_{c2}(3930)$ we use PDG 2021 values $M_{R}=3922.2 \pm 1.0$ MeV, $\Gamma_R=35.3 \pm 2.8$ \cite{ParticleDataGroup:2020ssz}, the couplings $g_{R\gamma\gamma}$ $g_{R D \bar{D}}$ cannot be fixed due to unknown branching fractions and will be absorbed into the unknown normalisation parameter (see Sec. \ref{subsec:ggdd}).

\begin{table*}[t]
\renewcommand*{\arraystretch}{1.3}
\begin{tabular*}{\textwidth}[t]{@{\extracolsep{\fill}}l|lllccll@{}}
\hline\hline
& $C_0$ & $C_1$ & $C_2$ & $N_2/N_0 \times 10^2$ & $\bar{\chi}^2_\text{comb}$  & $\bar{\chi}^2_\text{c}$  & $\bar{\chi}^2_\text{n}$\\ 
\hline\hline 
Fit I to combined Belle data & -64.5(16.1) & 167.7(18.9) & - & 2.9(0.9) & 0.91 & 9.84 & 2.88 \\
Fit II to $D^{+}D^{-}, D^{0}\bar{D}^{0}$ Belle data & 888.1(16.0) &-2315.1(0.5) & 1613.5(11.9) & 1.3(0.4) & 1.08 & 0.96 & 0.98 \\
Fit III to $D^{+}D^{-}, D^{0}\bar{D}^{0}$ BaBar data& 996.3(103.8) &-2336.1(208.4) & 1552.6(118.1) & 0.6(0.2) & 3.29 & 2.26 & 3.24 \\
\hline\hline
\end{tabular*}
\caption{Fit parameters entering Eqs.~(\ref{ConfExpansion}) and (\ref{N0N2}). Fit I is a fit to a combined $\sigma_c(s) + \sigma_n(s)$ data from the Belle Collaboration \cite{Uehara:2005qd}, while Fit II and Fit III are the fits to the charged and neutral channel data from Belle \cite{Uehara:2005qd} and BaBar Collaborations \cite{Aubert:2010ab}, respectively. The individual $\bar{\chi}^2_\text{comb/c/n}\equiv\chi^2_\text{comb/c/n}/\text{d.o.f.}$ show how good each fit describe the combined, charged or neutral data-sets.\label{tab:results}}
\end{table*}

\section{Results and discussion}\label{sec:results}
\subsection{Experimental input}\label{subsec:exp}
Before implementing the dispersive approach, we would like to comment on the quality of data that serves as an input to our analysis. The statistics in both Belle \cite{Uehara:2005qd} and BaBar \cite{Aubert:2010ab} $\gamma\gamma \to D\bar{D}$ experiments are relatively low, and therefore the sum of charged and neutral production modes was presented as the main result. In this way, the interference between $I=0$ and $I=1$ contributions cancels out and since the $I=1$ amplitude is expected to be smooth, it is natural to associate any structure in the combined data with the $I=0$ resonances. However, it will become apparent that the separate treatment of the neutral and charged channels is necessary to obtain the correct result for the $I=0$, $D\bar{D}$ dynamics. Since the Born term contribution enters the $D^0\bar{D}^0$ channel only via rescattering, one can expect more events in the $D^+D^-$ channel, which is not the case for the data on hand. In \cite{Wang:2020elp} this discrepancy was attributed to the fact that more decay modes were analyzed for the neutral channel in both experiments and the additional artificial factor of 1/3 was included to compensate for it. We, however, refrain from making any assumptions regarding the nature of the difference and proceed with the given data in a standard way.

While the Belle data \cite{Uehara:2005qd} is not efficiency corrected, the efficiency decreases by only 10\% for the region of invariant mass between 3.8 and 4.2 GeV, and therefore, this effect is expected to be negligible considering the resolution of the data itself. In addition, the data is provided in terms of the events distribution and to compare it with the cross sections, an additional normalization factor has to be introduced as a fitting parameter. This fact limits the possibility to extract the meaningful two-photon couplings of the resonances or bound states.  Even though BaBar Collaboration provides efficiency corrected data, it is given only for the sum of neutral and charged channels. Since the information in each channel separately is essential for our analysis, we opt to use the non-efficiency corrected version of the data in each channel, which, however, suffers from even lower resolution.

The Belle Collaboration \cite{Chilikin:2017evr} data for the invariant mass distribution of the  $e^+e^-\to J/\psi D\bar{D}$ reaction is also problematic. First, it is not acceptance corrected and hence the results should be taken with caution. Second, the resolution of this data is even  poorer  than for the $\gamma\gamma \to D\bar{D}$ case. The binning of 50 MeV does not allow to separate the narrow $J^{PC}=2^{++}$ resonance $\chi_{c2}(3930)$, clearly seen in other experiments. This problem, however, can be circumvented by excluding one data point at $\sim 3930$ MeV. Aiming to analyze the most relevant part of this data close to the threshold, in total there are only five data points left, all with relatively large uncertainties.

\subsection{Analysis of the \ggdd process}\label{subsec:ggdd}
In the analysis of the $\gamma\gamma \to D\bar{D}$ data, we limit ourselves to the region below 4.0 GeV, where the leading contribution is coming from the S and D-wave amplitudes. The cross-sections for individual partial waves in charged ($c$) or neutral ($n$) channels are given by
\begin{equation}
	\sigma_{c/n,\lambda_1\lambda_2}^{(J)} (s)= (2J+1)\,\frac{\beta(s)}{32\,\pi\,s}\,|h^{(J)}_{c/n, \lambda_1\lambda_2}(s)|^2\,,
\end{equation}
where the following relation between the isospin and particle basis holds
\begin{align}\label{isospin}
	h^{(J)}_{c,\lambda_1\lambda_2}(s)&=-\frac{1}{\sqrt{2}}\,(h^{(J)}_{0,\lambda_1\lambda_2}(s)+h^{(J)}_{1,\lambda_1\lambda_2}(s))\, , \nonumber\\
	h^{(J)}_{n,\lambda_1\lambda_2}(s) &= -\frac{1}{\sqrt{2}}\,(h^{(J)}_{0,\lambda_1\lambda_2}(s) - h^{(J)}_{1,\lambda_1\lambda_2}(s))\,.
\end{align}
As it was mentioned in Sec. \ref{subsec:exp}, to fit \ggdd data we need to introduce the normalization factor to convert the theoretical cross-section to the number of events from the experimental plot and a factor $N_2$, which accounts for the $\chi_{c2}(3930)$ couplings. The total cross-section for the charged or neutral channels is then given by
\begin{equation}\label{N0N2}
	\sigma_{c/n}(s) \approx N_0\,\sigma_{c/n,++}^{(0)}(s)+N_2\,\sigma_{c/n,+-}^{(2)}(s)\,,
\end{equation}
where we neglected the helicity-0 component of the D-wave. In addition to the free parameters $N_0, N_2$, there are also coefficients of the conformal expansion (\ref{ConfExpansion}), which determine the form of the left-hand cuts in (\ref{DR_1}) and have to be fitted to the data. Apart from the standard $\chi^2$ criteria, their number is chosen in a way to ensure that the series (\ref{ConfExpansion}) converges in the physical region. The statistical uncertainties are then propagated using the parametric bootstrap technique for all parameters and derived quantities like pole positions.

\begin{figure*}[!t]
\centering
\includegraphics[width =0.45\textwidth]{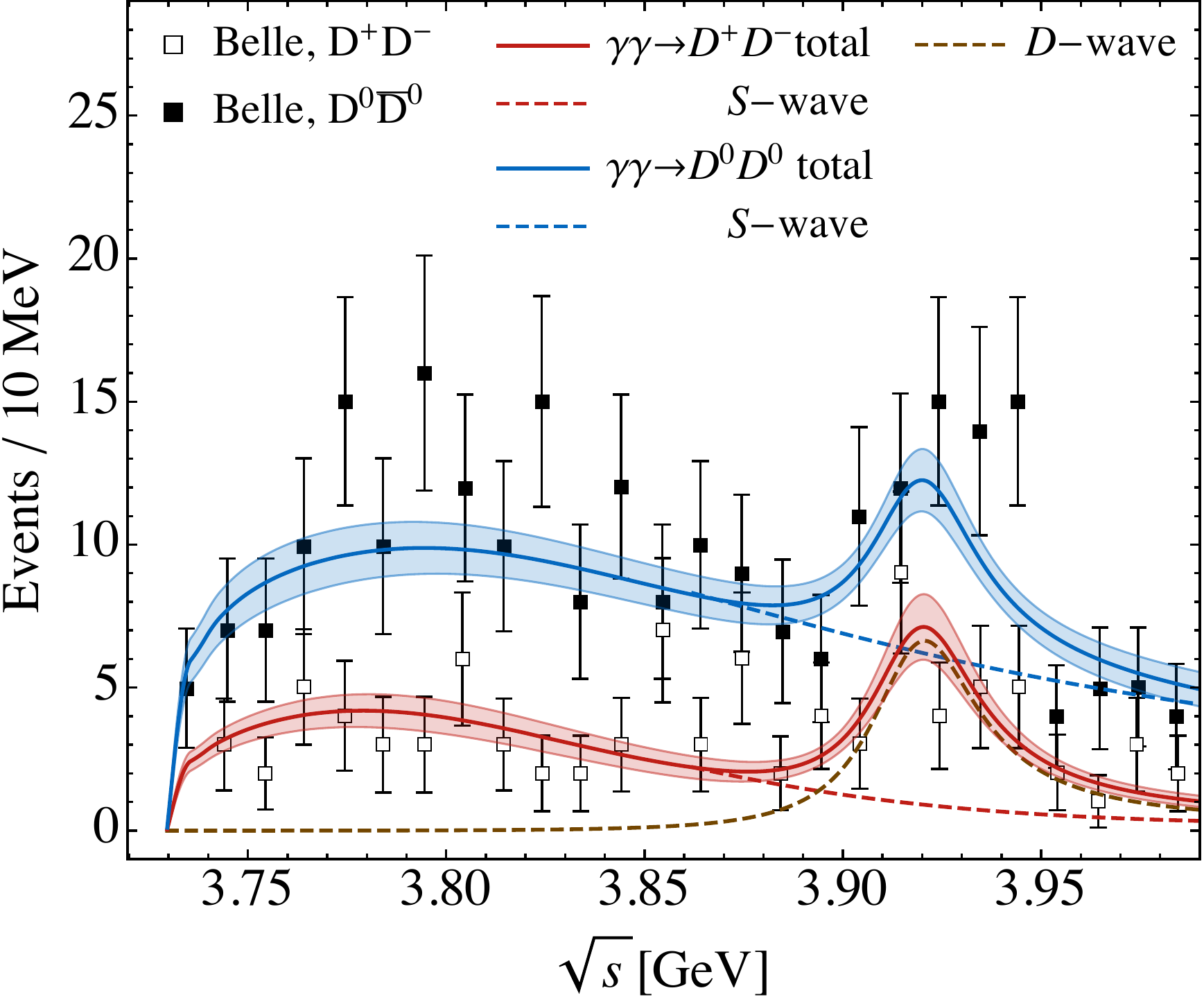}\qquad
\includegraphics[width =0.45\textwidth]{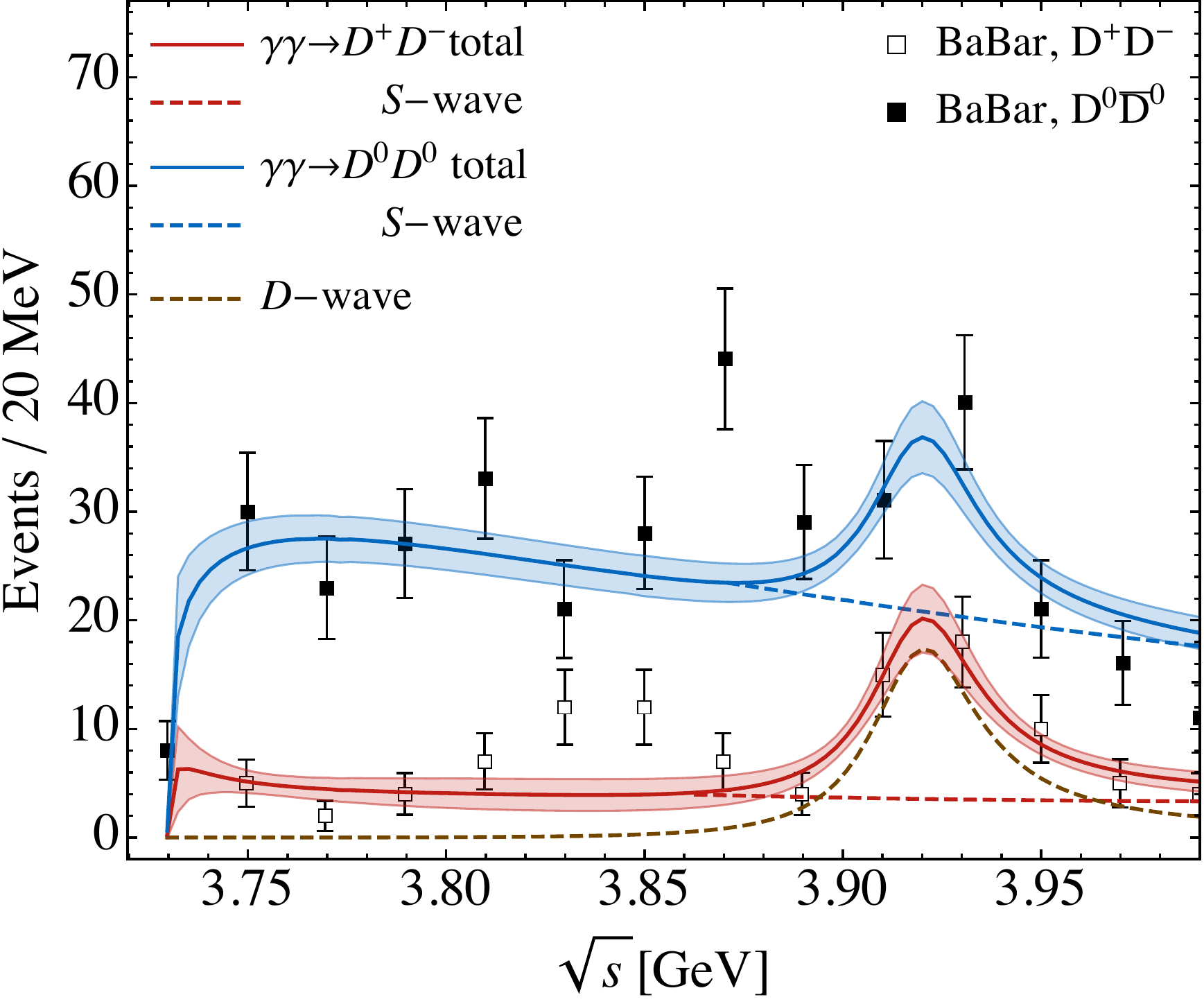}\\
\includegraphics[width =0.45\textwidth]{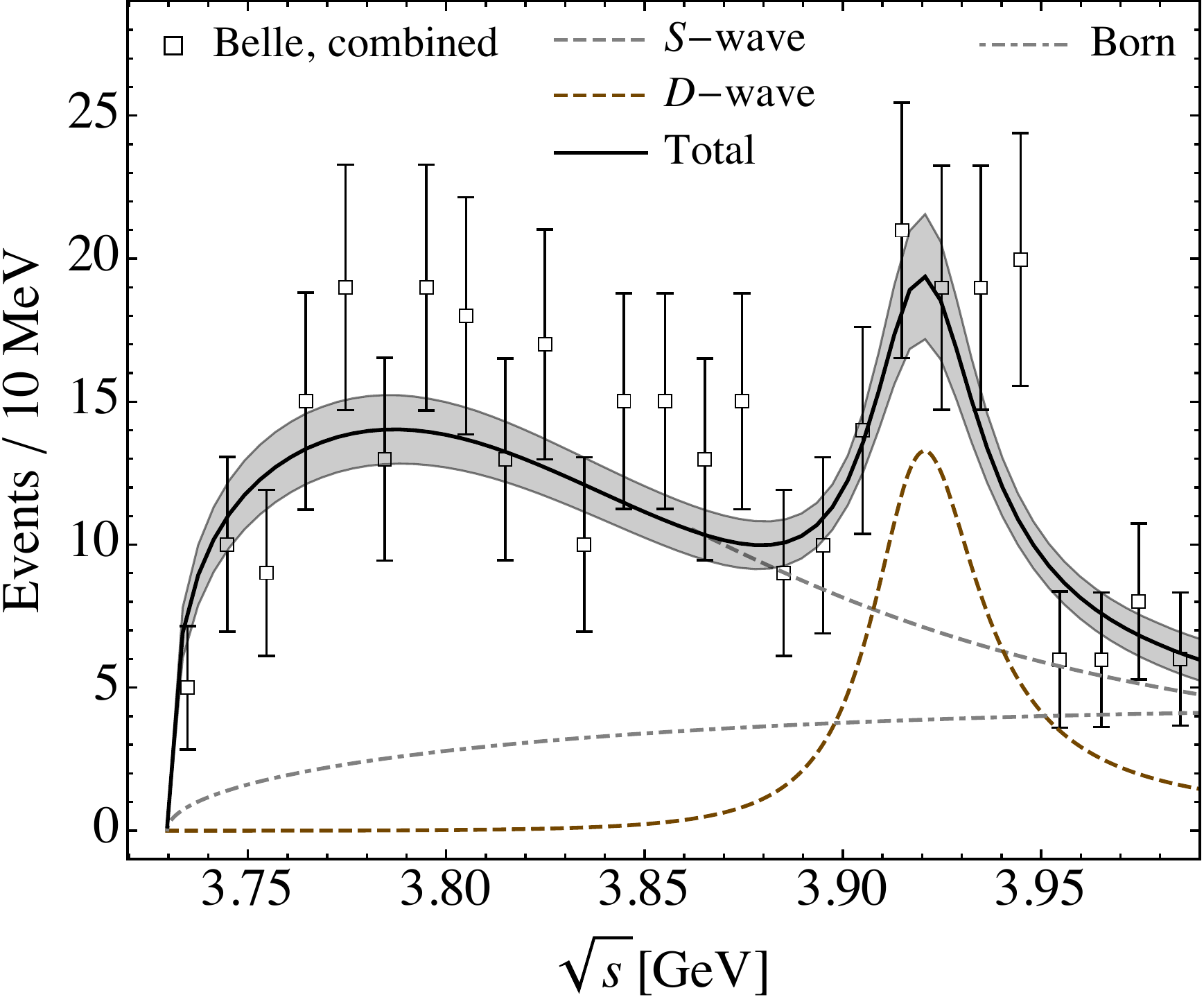}\qquad
\includegraphics[width =0.45\textwidth]{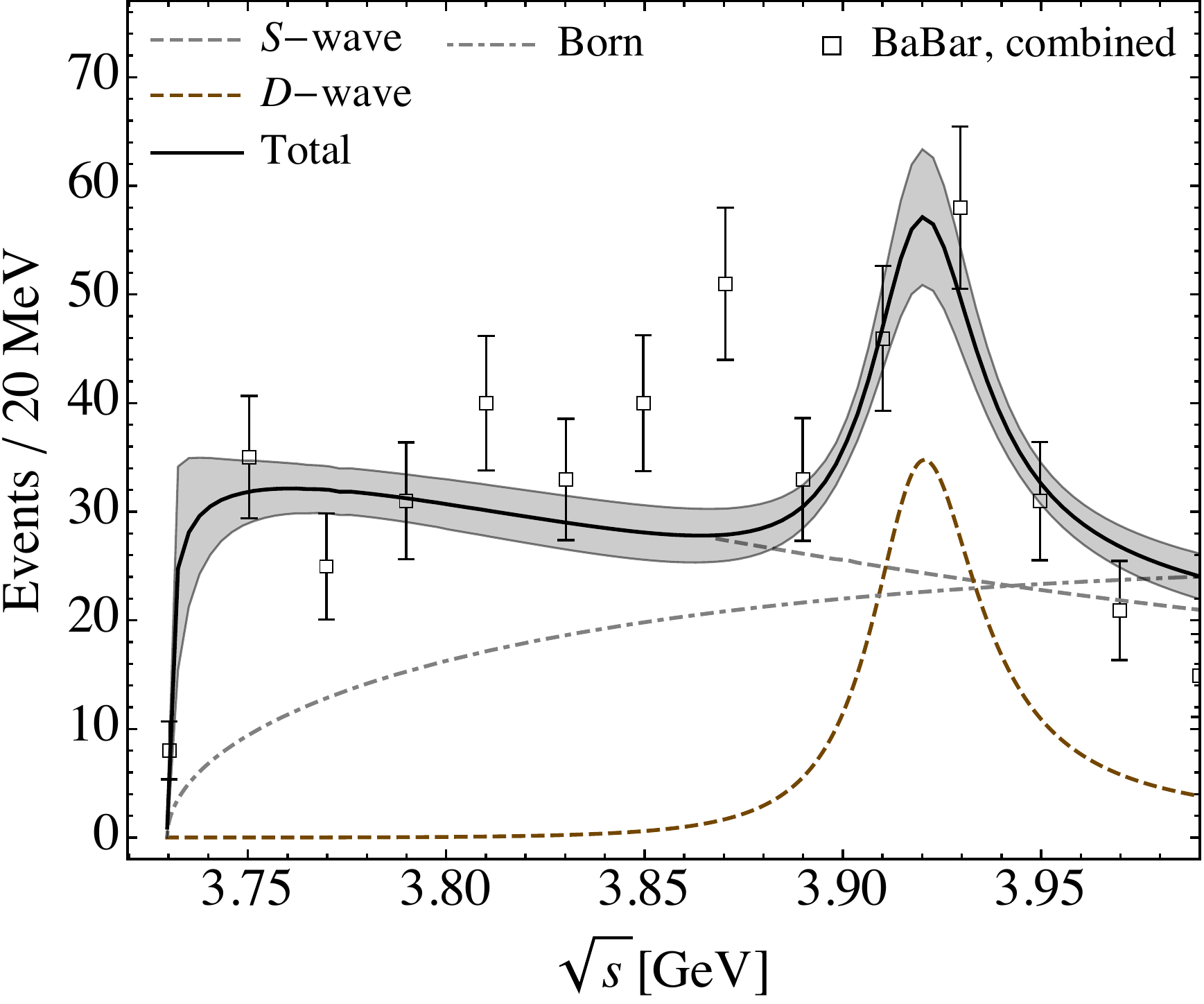}
\caption{Fit results compared to the data from the Belle \cite{Uehara:2005qd} (left panels) and BaBar \cite{Aubert:2010ab} (right panels) Collaborations. The top panels show the charged and neutral cross sections. For comparison, in the bottom panels the sum of two are presented in both cases.\label{fig:Belle}}
\end{figure*}
	
As the first step, we consider the combined \ggdd data
\begin{equation}
\sigma_{c}(s) + \sigma_{n}(s)
\end{equation}
from the Belle Collaboration \cite{Uehara:2005qd} alone as an input, similarly as it was done in \cite{Guo:2012tv}. We find that the fit to this data is already saturated with only two conformal expansion parameters and $\chi_\text{comb}^2/\text{d.o.f.} = 0.91$. Their values are listed in Table \ref{tab:results} (Fit I). For this fit we find a pole located at the second Riemann sheet with $\sqrt{s_P} = 3765.3(11.4) - i\,57.3(9.5)$ MeV, which is around $100$ MeV lower than the estimated values for the $X(3860)$ in \cite{Guo:2012tv} and significantly narrower. However, these results can not be directly compared, as the parameterization used in \cite{Guo:2012tv} does not establish the pole position in the complex $s$-plane, and only the mass and the width of the Breit-Wigner resonance are given. However, this fit can be misleading,
since it may not describe charged and neutral channels separately. In order to include this additional information, apart from the standard $\chi_\text{comb}^2/\text{d.o.f.}$ we introduce $\chi^2_\text{c}/\text{d.o.f.}$ and $\chi_\text{n}^2/\text{d.o.f.}$ tests, describing how well the given set of parameters reproduce the data in charged and neutral channels, respectively. We find, that even though Fit I accurately describes the combined data, it completely fails to account for the differences in separated data sets with $\chi^2_\text{c}/\text{d.o.f.} = 9.84$ and $\chi_\text{n}^2/\text{d.o.f.} = 2.88$.

\begin{figure}
    \centering
    \includegraphics[width =0.45\textwidth]{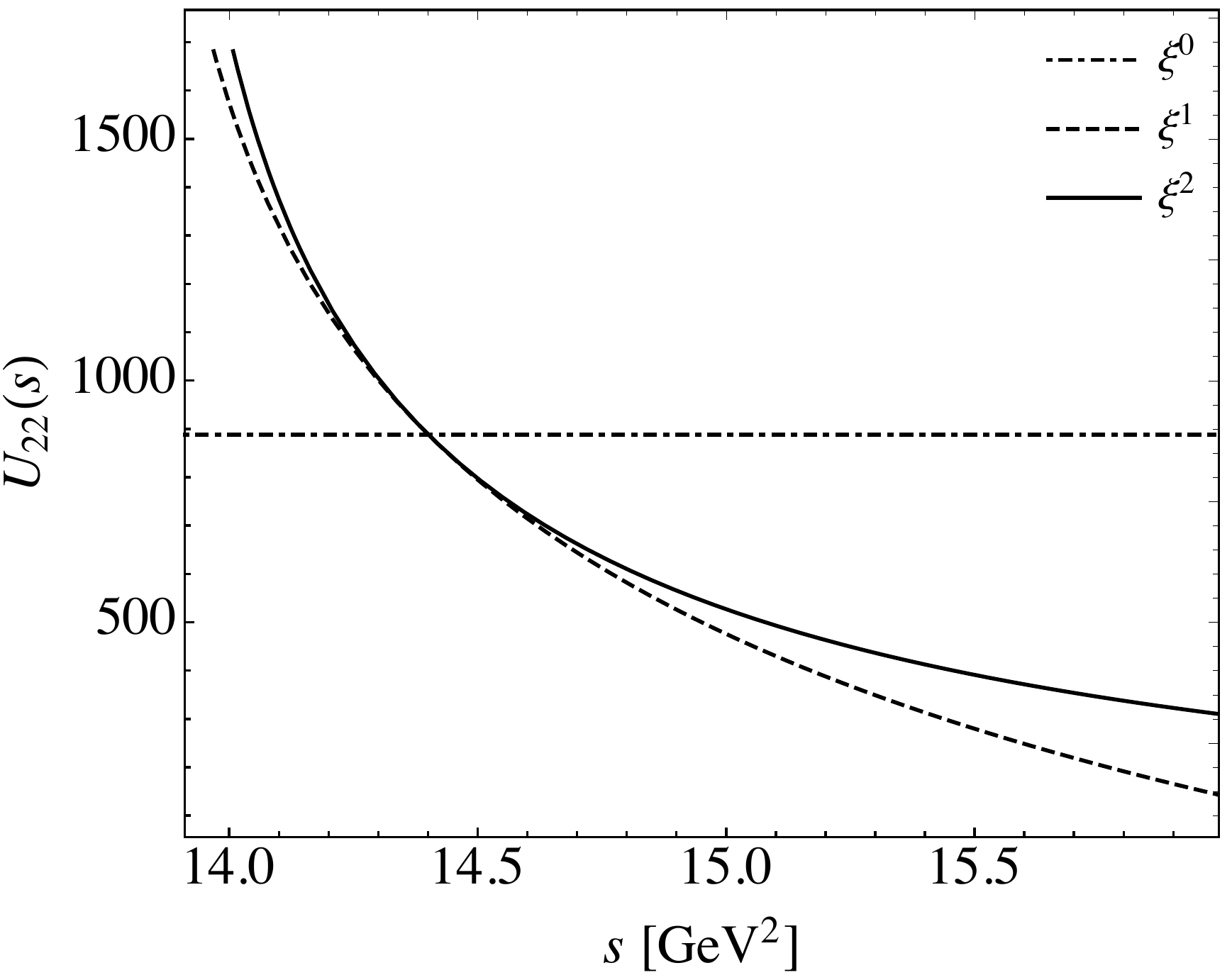}
    \caption{Convergence of the conformal expansion in Eq.~(\ref{ConfExpansion}) for $C_i$ values from Fit II.}
    \label{fig:convergence}
\end{figure}

\begin{figure*}[!t]
\centering
\includegraphics[width =0.45\textwidth]{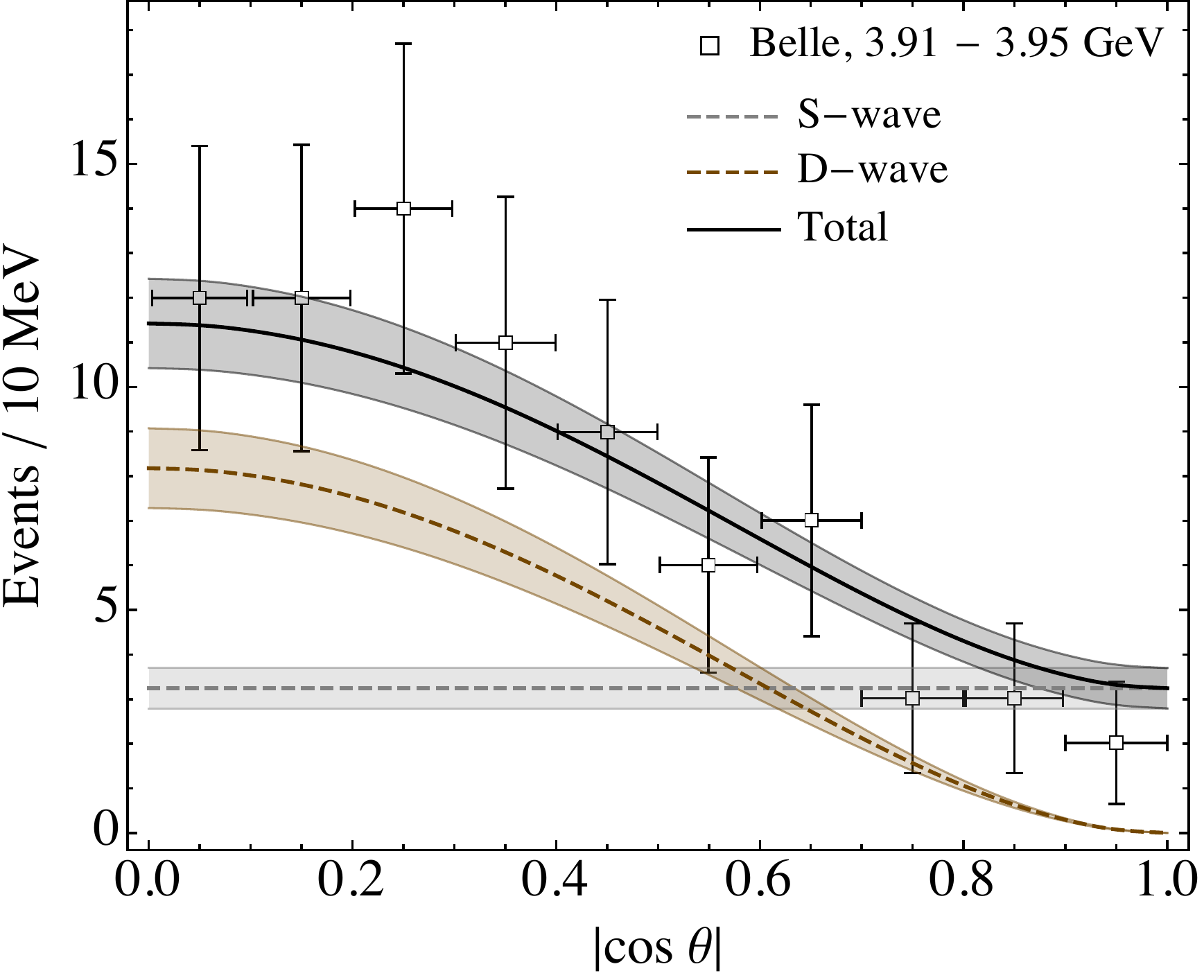}\qquad
\includegraphics[width =0.45\textwidth]{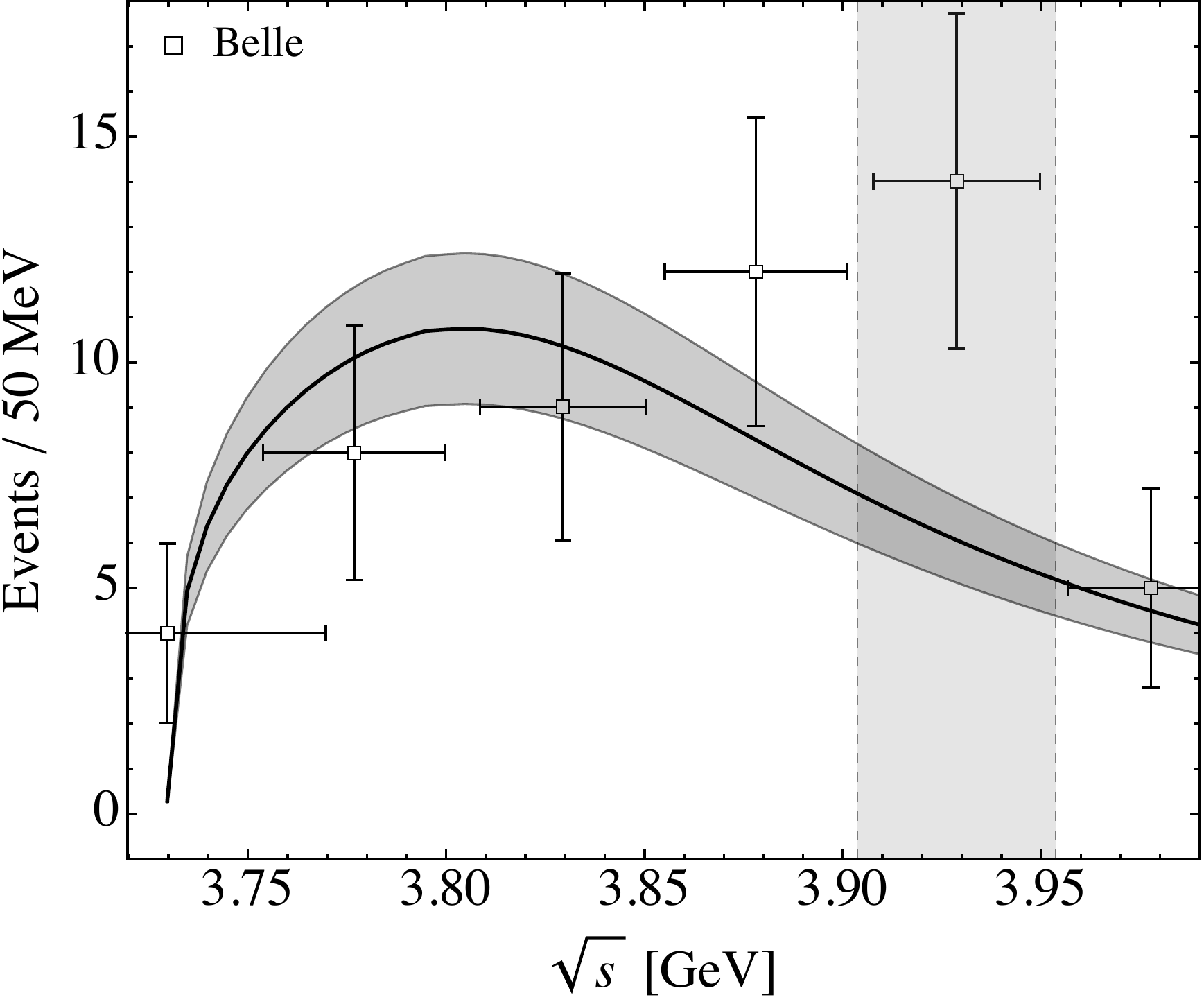}
\caption{Left: Angular distribution of the combined $\ggdd$ data from the Belle Collaboration \cite{Uehara:2005qd} in the energy range $3.91 - 3.95$ GeV compared to the theoretical curve calculated using the Fit II parameters. Right: The invariant $D\bar{D}$ mass distribution of the $e^+e^-\to J/\psi D\bar{D}$ process measured by the Belle Collaboration \cite{Chilikin:2017evr} compared to the S-wave dispersive result calculated using the Fit II parameters. The gray area covers the region where we expect a significant $\chi_{c2}(3930)$ contribution.\label{fig:jpsi} }
\end{figure*}

As a natural continuation, we perform a fit aiming to describe neutral and charged channels simultaneously. The best results are obtained with three conformal expansion parameters leading to $\chi^2_\text{c}/\text{d.o.f.} = 0.96$ and $\chi_\text{n}^2/\text{d.o.f.} = 0.98$. If compared to the combined data, this fit gives $\chi_\text{comb}^2/\text{d.o.f.} = 1.08$. The values of parameters are collected in Table \ref{tab:results} (Fit II) and the resulting curves are shown in Fig.~\ref{fig:Belle}. For this fit, instead of the pole in the complex plane, we find a bound state located under the $D\bar{D}$ threshold at
\begin{equation}
\sqrt{s_B}=3695(4)\,\text{MeV}\, .
\end{equation}
This bound state is stable against the variation of the $s_E$ parameter leading to negligible systematic uncertainties. We also checked explicitly that adding one more term in the conformal expansion barely changes the $\chi^2$. Note that even though the convergence of $U_{22}$ (see Fig.~\ref{fig:convergence}) and consequently $N_{22}$ is limited to energies $s>s_{L}$, the applicability domain of $D_{22}$  (which does not have a left-hand cut) is the whole complex plane and Eq.~(\ref{Eq:D=0}) is valid for energies sufficiently lower than the threshold.

From the Fit I we found, that fitting the combined data can lead to wrong results. Therefore, we do not consider the combined dataset of the BaBar data \cite{Aubert:2010ab}, which is efficiency corrected. Instead, we perform an auxiliary fit to the charged and neutral channels, which are not efficiency corrected, to show that even in case of very low data resolution we are able to obtain the aforementioned bound state with $\sqrt{s_\text{B}} = 3669.4(18.0)$ MeV. The resulting parameters are again tabulated in Table \ref{tab:results} (Fit III) and shown in Fig. \ref{fig:Belle}.

Regarding the energy region around $\chi_{c2}(3930)$ resonance, all fits provide similar results. While the resonance structure itself is governed by the Breit-Wigner-like parametrization (\ref{pw_D}), the height of the peak is partially defined by the tail of the $S$-wave contribution. This interplay between $S$- and $D$-waves can be studied on the level of angular distribution, for which the data from Belle Collaboration is provided in the region $\sqrt{s}=[3.91 - 3.95]$ GeV. Adopting the parameters of Fit II, we find a good agreement with the data (see Fig.~\ref{fig:jpsi}), showing that for these energies the angular distribution has a characteristic $D$-wave behavior with a constant shift from the $S$-wave contribution. Given the quality of the data, one cannot exclude that in the $\chi_{c2}(3930)$ region there is an additional small $S$-wave contribution from $\chi_{c0}(3930)$, which was recently claimed by the LHCb Collaboration \cite{LHCb:2020bls,*LHCb:2020pxc} in the $B^+\to D^+D^-K^+$ decays. However, we refrain from including it (as opposed to \cite{Chen:2012wy}), since the $\ggdd$ data will not be able to constrain it.

\subsection{Analysis of the $e^+ e^- \to \jp D \bar{D}$ process}\label{subsec:jpsi}

By considering only S-wave rescattering in the $D\bar{D}$ channel, the differential cross-section for the process $e^+e^-\to J/\psi D\bar{D}$ can be written as
\begin{equation}\label{Eq:eeJpsiDD}
	\frac{d\sigma}{d\sqrt{s}} = N\,\frac{\lambda^{1/2}(s,q^2,m^2_{J/\psi})\,\lambda^{1/2}(s,m^2_D,m^2_D)}{q^6\sqrt{s}}\left|D_{22}^{-1}(s)\right|^2\,,
\end{equation}
where $q$ is the $e^+e^-$ center of mass energy and the K\"allen function is defined by $\lambda(x,y,z)\equiv x^2+y^2+z^2-2(xy+xz+yz)$. In Eq.(\ref{Eq:eeJpsiDD}), similar to Eq.(\ref{AMPD}), the $D\bar{D}$ final state interaction is accounted for through the $D_{22}$ function. In this case, however, we use a simple model which only preserves unitarity in the direct $s$-channel and neglected possible contributions from the crossed channels (i.e. left-hand cuts). The latter are typically suppressed for the three body decays, but at the same require solving a set of Khuri-Treiman-type equations \cite{Niecknig:2012sj,*Danilkin:2014cra,*Albaladejo:2020smb,Guo:2016wsi,*Guo:2015zqa}. This study goes far beyond the scope of this paper and requires precise Dalitz plot data. 

With the limitations listed in Sec.~\ref{subsec:exp}, i.e. only a few available experimental points in the near-threshold region, the data from \cite{Chilikin:2017evr} alone is not constraining enough to provide a unique and meaningful solution without introducing additional assumptions. The same observation has been made in \cite{Wang:2019evy}. Therefore, we only check the consistency with the $\gamma\gamma\to D\bar{D}$ results by taking the best set of conformal expansion parameters given by Fit II and adjusting only the normalisation constant $N$ in Eq.~(\ref{Eq:eeJpsiDD}). Note, that we excluded the point $\sqrt{s}\sim 3930$ MeV, where we expect a significant $\chi_{c2}(3930)$ contribution. The results for the invariant mass distribution are shown in Fig. \ref{fig:jpsi}, where we choose the value of the $e^+e^-$ c.m. energy in the middle of the experimental region $9.46 - 10.87\,\text{GeV}$. The data is described with $\chi^2/\text{d.o.f} = 1.57$, indicating a very good agreement.

\subsection{Analogy to the $\gamma\gamma\to K\bar{K}$ process and $f_0(980)$}\label{subsec:ggkk}
	
\begin{figure}[t]
	\centering
	\minipage{0.45\textwidth}
	\includegraphics[width=\linewidth]{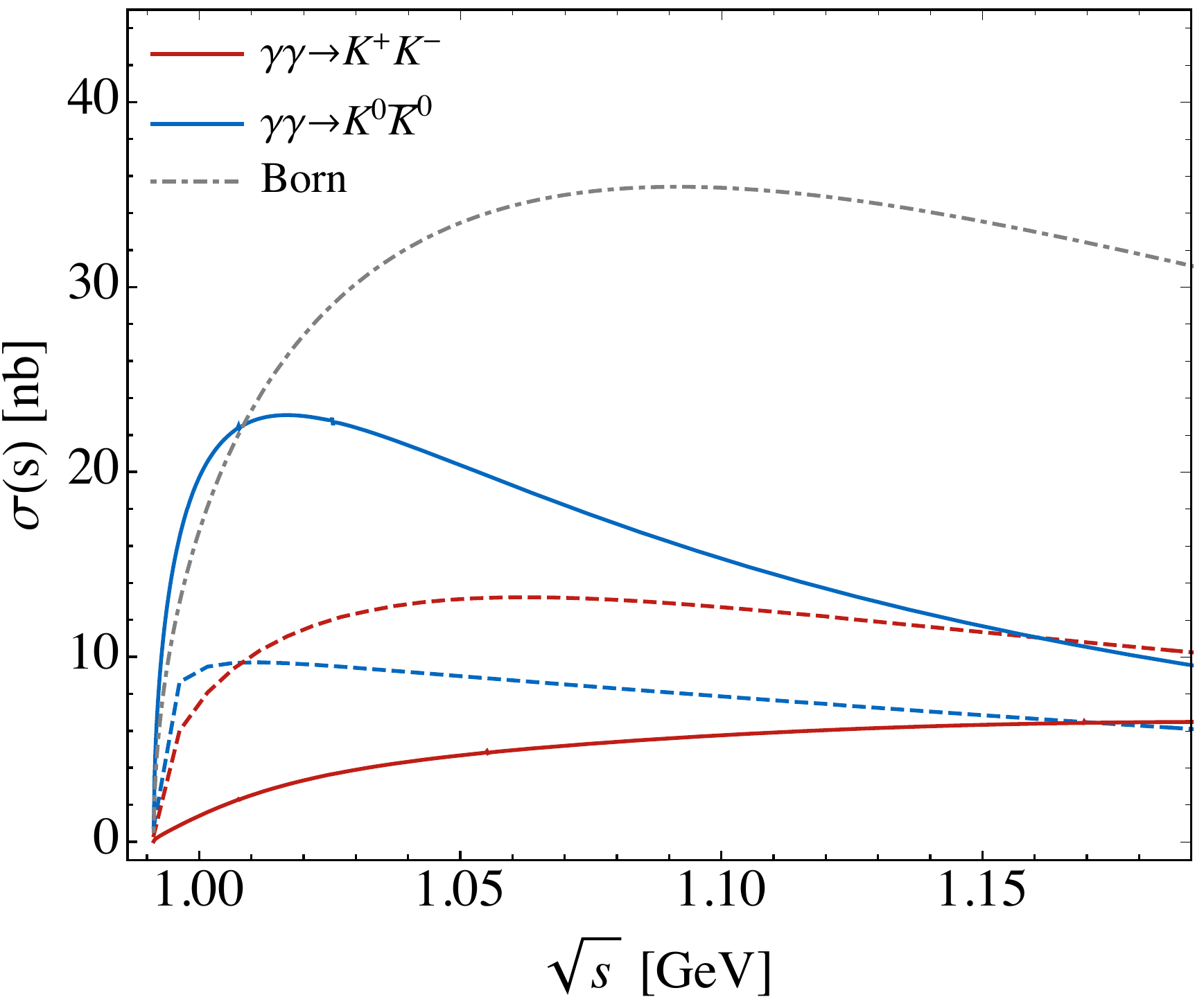}
	\endminipage\hfill
	\caption{The S-wave cross-sections for the reactions $\gamma\gamma\to K^+K^-$ (red) and $\gamma\gamma\to K^0\bar{K}^0$ (blue) under the assumption that $I=1$ contribution is dominated by the Born terms. Solid curves correspond to the hypothetical situation when there is no coupling to the $\pi\pi$ channel  and $f_0(980)$ is a pure bound state just below $K\bar{K}$ threshold, while dashed curves come from the $I=0$ $\{\gamma\gamma,\pi\pi,K\bar{K}\}$ coupled-channel analysis \cite{Danilkin:2020pak}. The Born contribution is shown as a dot-dashed curve.\label{fig:ggKK}}
\end{figure}

It is instructive to compare the obtained results for the $\gamma\gamma\to D\bar{D}$ process with a relatively well-known case of $\gamma\gamma\to K\bar{K}$. In the low-lying isoscalar S-wave sector, there are two resonances: $\sigma/f_0(500)$ and $f_0(980)$. While $\sigma/f_0(500)$ is known to be connected almost exclusively to the pion sector, $f_0(980)$ is a quasi-bound $K\bar{K}$ state. If we eliminate the connection to the $\pi\pi$ channel in the coupled-channel $\{\pi\pi,K\bar{K}\}$ dispersive analysis of \cite{Danilkin:2020pak}, then $f_0(980)$ resonance originally located at $\sqrt{s_p}=993(2)^{+2}_{-1}- i\,21(3)^{+2}_{-4}\,\rm{MeV}$ becomes a pure $K\bar{K}$ bound state with a binding energy of $\sqrt{s_B}=961\,\text{MeV}$. A similar feature was also observed in unitarized chiral perturbation theory calculations, see for instance \cite{Oller:1997ti}. On the level of cross-sections, if we treat $\gamma\gamma \to K\bar{K}\,(I=1)$ case on the same footing as the $\gamma\gamma \to D\bar{D}\,(I=1)$ process by taking only the Born terms\footnote{In the "real" world $\gamma\gamma \to K\bar{K}\,(I=1)$ channel has also a significant contribution from the $a_0(980)$ resonance through the $\{\gamma\gamma,\pi\eta,K\bar{K}\}$ coupled channels.}, then we observe a very similar pattern (compare Fig.~\ref{fig:ggKK} with upper panels of Fig.~\ref{fig:Belle}). While in the neutral channel the Born terms enter only via rescattering, it shows up stronger than the charged channel, due to destructive interference of the rescattering contribution with a pure Born amplitude at the level of the cross-section.

Similar to $f_0(980)$, one can also expect that the bound state $\sqrt{s_\text{B}} =3695(4)$ MeV found in the single-channel $\{D\bar{D}\}$ approximation will become a pole on the unphysical Riemann sheet once the channels $\{\pi\pi,K\bar{K},\eta\eta, ...\}$ will be switched on. However, the couplings to these channels are expected to be strongly suppressed due to their distant location \cite{Gamermann:2006nm}.

\section{Conclusion and outlook}\label{sec:conc}
In this work, we presented a theoretical analysis of  the reaction $\gamma\gamma\to D\bar{D}$  from threshold up to $4.0$ GeV. In order to account for the $D\bar{D}$ rescattering in the S-wave, we used a partial wave dispersive representation, which implements constraints from analyticity and exact unitarity. The left-hand cut contributions were accounted for by performing a model independent conformal mapping expansion, whose coefficients were fitted to the experimental data. On top of the S-wave, the well-established narrow D-wave resonance $\chi_{c2}(3930)$ was taken into account explicitly in the $s$-channel.

In the analysis of the data from the Belle \cite{Belle:2005rte} and BaBar \cite{BaBar:2010jfn} Collaborations, we found that it is crucial to simultaneously describe both charged $\gamma\gamma\to D^+D^-$ and neutral $\gamma\gamma\to D^0\bar{D}^0$ channels. Within our approach, we found no broad resonance $X(3860)$ currently associated with $\chi_{c0}(2P)$ in PDG (2021) \cite{ParticleDataGroup:2020ssz}. Instead, we found a bound state, located below $D\bar{D}$ threshold at $\sqrt{s_B} = 3695(4)$ MeV. The dataset for the invariant $D\bar{D}$ mass distribution of the $e^+e^-\to J/\psi D\bar{D}$ reaction, in which the $X(3860)$ resonance was reported \cite{Chilikin:2017evr}, confirms the consistency of our results. Using the S-wave $D\bar{D}$ final state interaction, we described the $e^+e^-\to J/\psi D\bar{D}$ process reasonably well, by adjusting only the overall normalization.

The bound state in the dispersive analysis without CDD poles qualifies for a molecular state. It is also consistent with other theoretical predictions \cite{Gamermann:2006nm, Prelovsek:2020eiw,Nieves:2012tt,*Hidalgo-Duque:2012rqv, *Hidalgo-Duque:2013pva, *Baru:2016iwj} and the absence of the broad near-threshold resonance was recently observed by experimental analysis by LHCb Collaboration \cite{LHCb:2020bls,*LHCb:2020pxc}. The detailed study of the properties of the found bound state $X(3695)$, however, requires more refined experimental input which can be achieved at Belle II. For this purpose it may be promising to search for the radiative decay $\psi(3770) \to \gamma X(3695)$, in analogy with $\psi(3770) \to \gamma \chi_{c0}$ radiative decay measurement at BESIII~\cite{BESIII:2015cby}. Furthermore, the existence such state $X(3695)$ may be tested in direct production at PANDA@FAIR.

\section*{Acknowledgements}
This work was supported by the Deutsche Forschungsgemeinschaft (DFG, German Research Foundation), in part through the Collaborative Research Center [The Low-Energy Frontier of the Standard Model, Projektnummer 204404729 - SFB 1044], and in part through the Cluster of Excellence [Precision Physics, Fundamental Interactions, and Structure of Matter] (PRISMA+ EXC 2118/1) within the German Excellence Strategy (Project ID 39083149). O.D. acknowledges funding by DAAD.
	
\bibliographystyle{apsrevM}
\bibliography{bibliography.bib}
	
\end{document}